\documentclass[prd,showkeys,floatfix,twocolumn,amsmath,amssymb,floatfix]{revtex4}
\usepackage{graphicx}
\usepackage{multirow}
\usepackage{subfigure}
\usepackage[colorlinks,citecolor=blue,anchorcolor=red,menucolor=red,linkcolor=red,filecolor=red,runcolor=red,urlcolor=blue,frenchlinks=red]{hyperref}
\usepackage{enumitem}
\usepackage{slashed}
\usepackage[normalem]{ulem}

\makeatletter
\renewcommand{\@thesubfigure}{\hskip\subfiglabelskip}
\makeatother
\allowdisplaybreaks[3]

\begin{document}
%=====================================================================================
\title{Interpretation of $\Omega(2012)$ as a $\Xi(1530)\bar{K}$ molecular state}
%=====================================================================================
\author{Xiang Yu$^{1}$}
\author{Jin-Peng Zhang$^1$}
\author{Xu-Liang Chen$^1$}
\author{Ding-Kun Lian$^2$}
\author{Qi-Nan Wang$^3$}
\author{Wei Chen$^{1,4}$}
\email{chenwei29@mail.sysu.edu.cn}

\affiliation{$^1$School of Physics, Sun Yat-sen University, Guangzhou 510275, China\\
$^2$School of Physics, Southeast University, Nanjing 210094, China\\
$^3$College of Physical Science and Technology, Bohai University, Jinzhou 121013, China\\
$^4$Southern Center for Nuclear-Science Theory (SCNT), Institute of Modern Physics,
Chinese Academy of Sciences, Huizhou 516000, Guangdong Province, China
}

\begin{abstract}
We investigate the mass and strong decay properties of the $\Omega(2012)$ resonance using QCD sum rules, assuming it to be an S-wave $\Xi(1530)\bar{K}$ molecular pentaquark state with $I(J^{P})= 0(\frac{3}{2}^{-})$. A unified interpolating current is constructed, and the two-point correlation functions and three-point functions are calculated up to dimension-13 and 10 condensate terms in the OPE series, respectively. The negative-parity contribution is isolated by employing parity-projected sum rules. The two-body strong decays to $\Xi^0 K^-$ and $\Xi^- \bar{K}^0$ are studied via their three-point correlation functions.  Our analysis yields a mass of $2.00 \pm 0.15~\mathrm{GeV}$ and a total two-body decay width of $\Gamma = 0.96^{+0.79}_{-0.41}~\mathrm{MeV}$ for the $\Xi(1530)\bar{K}$ molecular state. The ratio of the two-body decay branching fractions is obtained as $\mathcal{R}^{\Xi^- \bar{K}^0}_{\Xi^0 K^-} = 0.85$. These results are compatible with the experimental data for the $\Omega(2012)$ within uncertainties and support its interpretation as a $\Xi(1530)\bar{K}$ molecular pentaquark state.
\end{abstract}

\maketitle

\section{Introduction}
\label{sec:intro}
Quantum chromodynamics (QCD) is the fundamental theory of the strong interaction. While conventional hadrons are well described as quark-antiquark mesons and three-quark baryons in the quark model~\cite{Gell-Mann:1964ewy,Zweig:1964jf}, QCD also permits the existence of exotic hadron states such as multiquark states, hybrid mesons,  glueballs and so on. Detailed discussions can be found in recent comprehensive reviews~\cite{Meyer:2015eta,Chen:2016qju,Clement:2016vnl,Guo:2017jvc,Liu:2019zoy,Brambilla:2019esw,Chen:2022asf,Liu:2024uxn,Esposito:2016noz,Meng:2022ozq,Lebed:2016hpi,Hosaka:2016pey,Ali:2017jda,Olsen:2017bmm,Wang:2025sic}.

In 2018, the Belle Collaboration first observed a narrow structure, denoted $\Omega(2012)$, in the $\Xi^{0}K^{-}$ and $\Xi^{-}K^{0}_{S}$ invariant mass spectra, providing the first experimental evidence for this excited $\Omega$ baryon~\cite{Belle:2018mqs}. Prior to this discovery, the mass spectrum of the $\Omega$ family had been extensively studied using various theoretical approaches~\cite{Aliev:2016jnp,An:2014lga,An:2013zoa,Bijker:2000gq,Bijker:2015gyk,Capstick:1986ter,Carlson:2000zr,Chao:1980em,Chen:2009de,Engel:2013ig,Faustov:2015eba,Goity:2003ab,Kalman:1982ut,CLQCD:2015bgi,Loring:2001ky,Matagne:2006zf,Oh:2007cr,Pervin:2007wa,Schat:2001xr}, many of which predicted a pair of negative-parity excited $\Omega^-$ states with $J^P = \frac{1}{2}^-$ and $J^P = \frac{3}{2}^-$ in the mass region near 2~GeV. The existence of the $\Omega(2012)$ was subsequently reinforced by Belle through its observation in weak decays of the $\Omega_c$ baryon, offering an independent production channel and further consolidating its resonance nature~\cite{Belle:2021gtf}. More recently, the ALICE Collaboration reported the discovery of $\Omega(2012)$ with a significance of $15\sigma$ in $pp$ collisions at the LHC~\cite{ALICE:2025atb}, while the BESIII Collaboration presented evidence for a new production mechanism of this state via the process $e^+e^-\rightarrow\Omega(2012)\bar\Omega^++c.c.$ with a significance of $3.5\sigma$~\cite{BESIII:2024eqk,Wang:2026daa}. In PDG~\cite{ParticleDataGroup:2024cfk}, the mass and total decay width of $\Omega(2012)$ are collected to be $m=2012.4\pm 0.9$ MeV and $\Gamma=6.4^{+3.0}_{-2.6}$ MeV, and the branching fraction ratio $\mathcal{R}^{\Xi^- \bar{K}^0}_{\Xi^0 K^-} = 0.83\pm 0.21$. 

The discovery of this multistrange resonance has stimulated considerable theoretical interest, which focused on two broad interpretive frameworks. One framework regards $\Omega(2012)$ as a conventional excited baryon. In particular, based on its relatively narrow width, it has been argued that the most likely spin-parity assignment is $J^{P}=\frac{3}{2}^{-}$ ~\cite{Aliev:2018syi,Aliev:2018yjo,Arifi:2022ntc,Hockley:2024aym,Liu:2019wdr,Menapara:2021vug,Polyakov:2018mow,Su:2024lzy,Wang:2022zja,Wang:2018hmi,Xiao:2018pwe,Zhong:2022cjx,Luo:2025cqs}. The alternative picture describes $\Omega(2012)$ as a molecular state dominated by the $\Xi^{*}\bar{K}$ configuration with $J^{P}=\frac{3}{2}^{-}$~\cite{Gutsche:2019eoh,Hu:2022pae,Huang:2018wth,Ikeno:2022jpe,Ikeno:2020vqv,Lin:2018nqd,Lin:2019tex,Liu:2020yen,Lu:2020ste,Pavao:2018xub,Valderrama:2018bmv,Xie:2024wbd,Zeng:2020och,Shen:2025xcq}. Within this molecular scenario, the generation of a $J^{P}=\frac{1}{2}^{-}$ state is generally disfavored. This interpretation is motivated by the mass of $\Omega(2012)$ lying close to the $\Xi(1530)\bar{K}$ threshold. Notably, both scenarios are capable of reproducing key experimental observables, such as the mass, narrow width, and production in $\Omega_{c}$ decays, suggesting that the physical $\Omega(2012)$ state may involve a nontrivial interplay between compact three-quark and hadron molecule components~\cite{Lu:2022puv,Zhong:2022cjx,Han:2025gkp}. A comprehensive overview of the experimental and theoretical status of $\Omega(2012)$ can be found in the recent reviews~\cite{Hyodo:2020czb,Xie:2024wbd}.

A crucial distinction between these competing interpretations arises from their predictions for decay patterns, particularly the three-body decay channels. Conventional three-quark models generally favor dominant two-body $\bar{K}\Xi$ decays, with the three-body $\bar{K}\pi\Xi$ mode being strongly suppressed. In the QCD sum rule approach with the conventional $sss$ configuration, the calculated mass is in good agreement with the experimental value~\cite{Su:2024lzy,Aliev:2018syi}, whereas the computed two-body decay width is already larger than the measured total width~\cite{Aliev:2018yjo}, implying very small branching ratio of the three-body decay channel. In contrast, molecular scenarios naturally allow for sizable contributions from the $\bar{K}\pi\Xi$ channel due to strong meson–baryon couplings. Experimentally, early analyzes reported no evidence for such three-body decays~\cite{Belle:2019zco}. However, subsequent measurements by Belle Collaboration revisited this issue and reported the experimental value $\mathcal{R}^{\Xi \pi \bar{K}}_{\Xi \bar{K}} = 0.99 \pm 0.26 \pm 0.06$~\cite{Belle:2022mrg}. More recently, the ALICE Collaboration reported the branching fraction for the two-body decay channel as $\mathcal{B}[\Omega^*\rightarrow \Xi \bar{K}]=0.62^{+0.27}_{-0.17}$ in 2025~\cite{ALICE:2025atb}. Assuming that all $\Omega(2012)$ decays proceed exclusively via either $\Omega(2012)\rightarrow \Xi \bar{K}$ or $\Omega(2012)\rightarrow \Xi\pi \bar{K}$, the ALICE result is consistent with the Belle measurement within the uncertainties.

In this paper, we employ QCD sum rules to carry out a systematic investigation of the mass and the two-body decay width of $\Omega(2012)$, treating it as an S-wave $\Xi(1530)\bar{K}$ molecular pentaquark state. Our calculations for both the mass and the two-body decay properties of the $\Omega(2012)$ are consistent with the available experimental results, supporting its interpretation as a $\Xi(1530)\bar{K}$ molecular pentaquark with $J^{P}=\frac{3}{2}^{-}$. Nevertheless, a definitive conclusion requires a further analysis of its three-body decay channels, which we leave for future work.

This paper is organized as follows. In Sec.~\ref{sec:current}, we introduce the interpolating currents and outline the QCD sum rules analysis. Sec.~\ref{sec:numerical} presents the numerical results of hadron mass and two-body strong decays. Finally, concluding remarks are provided in Sec.~\ref{sec:summary}.

\section{Two-point and three-point QCD sum rules}
\label{sec:current}
\subsection{The $\Xi(1530)\bar{K}$ molecular interpolating current with $J^{P}=3/2^{-}$}
In this subsection, we construct an interpolating current that describes a meson–baryon molecular state with $J^{P}=\frac{3}{2}^{-}$. The interpolating current for the $\bar{K}$ meson is given by
\begin{equation}
    J_{\bar{K}}(x)=\bar{q}^d(x)\mathrm{i}\gamma_5s^d(x)\, ,
    \label{currentK}
\end{equation}
while the current for the $\Xi(1530)$ ($\Xi^*$) baryon with $J^{P}=\frac{3}{2}^{+}$ reads
\begin{equation}
\begin{aligned}
    J_{\Xi^*}(x)=&\sqrt{\frac{1}{3}}[2s^{aT}(x)\mathcal{C}\gamma_{\mu}q^{b}(x)s^c(x)\\
    &+s^{aT}(x)\mathcal{C}\gamma_{\mu}s^{b}(x)q^c(x)]\, ,
\end{aligned}
\label{current1530}
\end{equation}
in which $a,b,c,d$ are color indices, and $\mathcal{C}=\mathrm{i}\gamma_2\gamma_0$, $s$ and $q$ are the strange and light ($u/d$) quark fields respectively. The current $J_{\Xi^*}$ has been widely adopted in QCD sum rules of decuplet baryons~\cite{Reinders:1982qg}. One shall find the currents $J_{K^-}$ and $J_{\Xi^{*0}}$ for $q=u$, while the currents $J_{\bar{K}^0}$ and $J_{\Xi^{*-}}$ for $q=d$ in Eqs.~\eqref{currentK}-\eqref{current1530}.
Then we can construct an S-wave $\Xi^*\bar K$ molecular pentaquark current with $J^{P}=\frac{3}{2}^{-}$ as
\begin{equation}
  \begin{aligned}
    \eta_{\mu}=&\sqrt{\frac{1}{3}}\epsilon^{abc}[\bar{q}^d(x)\mathrm{i}\gamma_5s^d(x)][2(s^{aT}(x)\mathcal{C}\gamma_{\mu}q^{b}(x))s^c(x)\\
    &+(s^{aT}(x)\mathcal{C}\gamma_{\mu}s^{b}(x))q^c(x)]\, .
  \end{aligned}
  \label{current2012}
\end{equation}

It is worth emphasizing that alternative pentaquark currents can also be constructed with the same quantum numbers but different Dirac structures, such as
\begin{equation}
    \begin{aligned}
        &\sqrt{\frac{1}{3}} \epsilon ^{abc} [\overline{q}^{d} (x)\mathrm{i}\gamma _{5} s^{d} (x)][2(s^{aT} (x)\mathcal{C}\gamma _{\mu } q^{b} (x))\sigma _{\mu \nu } s^{c} (x)\\
        &+(s^{aT} (x)\mathcal{C}\gamma _{\mu } s^{b} (x))\sigma _{\mu \nu } q^{c} (x)]
    \end{aligned}
    \label{current1}
\end{equation} 
or
\begin{equation}
    \begin{aligned}
        &\sqrt{\frac{1}{3}} \epsilon ^{abc} [\overline{q}^{d} (x)\mathrm{i}\gamma _{5} s^{d} (x)][2(s^{aT} (x)\mathcal{C}\sigma _{\mu \nu } q^{b} (x))\gamma _{\mu } s^{c} (x)\\
        &+(s^{aT} (x)\mathcal{C}\sigma _{\mu \nu } s^{b} (x))\gamma _{\mu } q^{c} (x)]\, .
    \end{aligned}
    \label{current2}
\end{equation}
A Fierz rearrangement, however, reveals that these currents are not independent but are equivalent to one another~\cite{Ioffe:1981kw}. It follows that, within this construction, Eq.~\eqref{current2012} represents the unique interpolating current for the $J^{P}=3/2^{-}$ $\Xi(1530)\bar{K}$ molecular state. 

Considering the isoscalar nature of $\Omega(2012)$, one can finally achieve the interpolating current coupling to this state as 
\begin{equation}
J_\mu(x)=\frac{1}{\sqrt{2}}\left(J_{\Xi^{*0}}\cdot J_{K^-}-J_{\Xi^{*-}}\cdot J_{\bar{K}^0}\right)\, . \label{currentomega2012}
\end{equation}
In the following analyses, we shall use this current to study the hadron mass and the strong decay properties of $\Omega(2012)$.

\subsection{Two-point correlation function}
The molecular interpolating current $J_{\mu}$ in Eq.~\eqref{currentomega2012} can couple to both the negative-parity and positive-parity states  through different coupling relations
\begin{align}
  \langle0|J_\mu|\Omega^*;3/2^-\rangle & =f_-u_\mu(p)\, , \\
  \langle0|J_\mu|\Omega^{*\prime};3/2^+\rangle & =f_+\gamma_5u_\mu(p)\, ,
  \label{couple2012n}
\end{align}
where $f_{\pm}$ are coupling constants and $u_\mu(p)$ is the Rarita-Schwinger vector-spinor. Thus, the two-point correlation function induced by such a current contains both information. It can be written as
\begin{equation}
  \begin{aligned}
    \Pi_{\mu\nu}(p^2)&\equiv \mathrm{i} \int \mathrm{d}^dx ~\mathrm{e}^{\mathrm{i} p \cdot x}\left\langle0|T[J_{\mu}(x)\bar{J}_{\nu}(0)]|0\right\rangle\\
    &=-g_{\mu\nu}\Pi_{3/2}(p^2)+...\, ,
  \end{aligned}
  \label{twofunction}
\end{equation}
where $\Pi_{3/2}(p^2)$ is the invariant function corresponding to the hadron states with $J^P=\frac{3}{2}^{\pm}$~\cite{Chung:1981cc,Jido:1996ia}. 
In general, the interpolating current in Eq.~\eqref{current2012} may also couple to non-resonant $\Xi(1530)\bar{K}$ scattering states with $J^{P}=3/2^{-}$, potentially contaminating the invariant function $\Pi_{3/2}(p^2)$ in Eq.~\eqref{twofunction} and thereby hindering the investigation of the $\Xi(1530)\bar{K}$ molecular state. However, previous studies have shown that non-resonant scattering states do not saturate the multiquark QCD sum rules, and their contributions are generally negligible within the standard sum rule treatment~\cite{Wang:2020cme,Wang:2019gal,Albuquerque:2021tqd,Wang:2020fuh}. A detailed discussion of scattering states in multiquark QCD sum rules can be found in the recent review~\cite{Wang:2025sic}. 

On the phenomenological side, by inserting a complete set of intermediate states, the correlation function can be expressed as
\begin{equation}
    \begin{aligned}
        \Pi_{\mu\nu}^{\mathrm{phe}}(p^2)&=-g_{\mu\nu}\left[(f_-)^2\frac{\slashed{p}+m_{-}}{m_{-}^2-p^2}+(f_+)^2\frac{\slashed{p}-m_{+}}{m_{+}^2-p^2}\right]+...\\
        &=-g_{\mu\nu}\left[\Pi_{\slashed{p}}^{\text{phe}}(p^2)\slashed{p}+\Pi^{\text{phe}}_I(p^2)\right]+...\, .
    \end{aligned}
    \label{twofunctionphe}
\end{equation}
The invariant function $\Pi(p^2)$ can be described by the dispersion relation
\begin{equation}
    \Pi(p^2)=(p^2)^n\int^{\infty}_{0}\frac{\rho(s)}{s^n (s-p^2)} \mathrm{d}s + \sum_{k=0}^{n-1}a_k (p^2)^k\, ,
    \label{dispersion}
\end{equation}
where $a_k$ is the subtraction constant and $\rho(s)=\frac{1}{\pi}\operatorname{Im} \Pi(s)$ is defined as the spectral function. According to Eqs.~\eqref{twofunctionphe}-\eqref{dispersion}, one finds the following two spectral functions 
\begin{equation}
\begin{aligned}
  \rho_{\slashed{p}}^{\mathrm{phen}}(s)&=f_-^2\delta(s-m_-^2)+f_+^2\delta(s-m_+^2)\, , \\
  \rho_{I}^{\mathrm{phen}}(s)&=f_-^2m_-\delta(s-m_-^2)-f_+^2m_+\delta(s-m_+^2)\, , \\
  \end{aligned}
  \label{rho1and0}
\end{equation}
from which we can separate the spectral densities for negative- and positive-parity states as
\begin{equation}
\label{rho+-}\rho^{\text{phe}}_{\mp}=\sqrt{s}\rho_{\slashed{p}}^{\text{phe}}(s)\pm\rho_{I}^{\text{phe}}(s)\, .
\end{equation}

At the quark-gluon level, the correlation function is computed using the operator product expansion (OPE), expressed in terms of quark masses and various QCD parameters. 
We adopt the $d$-dimensional coordinate-space expression for the light quark propagator~\cite{Zhang:2025fuz}
\begin{widetext}
  \begin{equation}
    \begin{aligned}
  S^{i j} (x) & =  \frac{\text{i} \Gamma \left( \frac{d}{2} \right)
  \slashed{x}}{2 \pi^{d / 2} (- x^2)^{d / 2}} \delta^{i j} + \frac{m \Gamma \left(
  \frac{d}{2} - 1 \right)}{4 \pi^{d / 2} (- x^2)^{d / 2 - 1}} \delta^{i j} -
  \frac{\delta^{i j}}{12} \langle \bar{\psi} \psi \rangle +
  \frac{\text{i} m \delta^{i j}}{12 d} \langle \bar{\psi} \psi \rangle
  \slashed{x} - \frac{\delta^{i j}}{48 d} \langle g \bar{\psi} \sigma G \psi
  \rangle x^2\\
  & -  \frac{\text{i} \delta^{i j} x^2 \slashed{x}}{2^4 3^4 (d + 2)} g^2 \langle
  \bar{\psi} \psi \rangle^2 + \frac{\text{i} \delta^{i j} m x^2
  \slashed{x}}{2^4 3 d (d + 2)} \langle g \bar{\psi} \sigma G \psi \rangle-
  \frac{\delta^{i j} x^4 \langle \bar{\psi} \psi \rangle \langle g^2 G^2
  \rangle}{2^6 3^2 d (d + 2)} - \frac{\delta^{i j} x^4}{2^4 3^4 d (d + 2)} g^2
  m \langle \bar{\psi} \psi \rangle^2\\
  & -  \frac{\text{i} \delta^{i j} \Gamma \left( \frac{d}{2} - 1 \right)
  \slashed{x} \langle g^3 f G^3 \rangle}{2^8 3^3 d (d + 2) \pi^{d / 2} (- x^2)^{d
  / 2 - 3}}+ \frac{\Gamma \left( \frac{d}{2} - 1 \right) \gamma^{\mu}
  \slashed{x} \gamma^{\nu}}{16 \pi^{d / 2} (- x^2)^{d / 2 - 1}} g G_{\mu \nu}^a
  T^a_{i j}\\
  & +  \left[ \frac{\Gamma \left( \frac{d}{2} - 2 \right)
  ({\gamma^{\mu \rho \nu} + \gamma^{\rho \mu \nu} - 4
  g^{\mu \rho} \gamma^{\nu}})}{96 \pi^{d / 2} (- x^2)^{d / 2 - 2}} + \frac{\Gamma \left( \frac{d}{2} - 1 \right) \left( x^{\mu} \gamma^{\rho}
  \slashed{x} \gamma^{\nu} + x^{\rho} \gamma^{\mu} \slashed{x} \gamma^{\nu}
  \right)}{48 \pi^{d / 2} (- x^2)^{d / 2 - 1}} \right] g G_{\mu \nu ; \rho}^a
  T^a_{i j}\\
  & +  \left( \frac{- \Gamma \left( \frac{d}{2} - 2 \right) \left(
  {2 g^{\{ \mu \rho } x^{ \sigma \}}}
  + g^{\{ \mu \rho } \gamma^{ \sigma \}} \slashed{x}
  \right)}{2^8 \times 3 \pi^{d / 2} (- x^2)^{d / 2 - 2}} + \frac{\Gamma \left(\frac{d}{2} - 1 \right) x^{\{ \mu} x^{\rho} \gamma^{\sigma \}} \slashed{x}}{192 \pi^{d / 2} (- x^2)^{d / 2 - 1}} \right)
  \gamma_{\nu} g G_{\mu \nu ; \rho \sigma}^a T^a_{i j}\\
  & +  g^2 G^a_{\mu \nu} G^b_{\rho \sigma} (T^a T^b)_{i j} \left[ \frac{-
  \text{i} \Gamma \left( \frac{d}{2} - 1 \right) x^{\nu} x^{\sigma}
  \gamma^{\mu} \slashed{x} \gamma^{\rho}}{96 \pi^{d / 2} (- x^2)^{d / 2 - 1}} \right.+{\frac{- \text{i} \Gamma \left( \frac{d}{2} - 2 \right)
  g^{\nu \sigma} \gamma^{\mu} \slashed{x} \gamma^{\rho}}{192 \pi^{d / 2} (-
  x^2)^{d / 2 - 2}}}  \\
  &+  \frac{ \text{i} \Gamma \left( \frac{d}{2} - 2 \right)}{2^8 \times 3
  \pi^{d / 2} (- x^2)^{d / 2 - 2}} ( - 6 g^{\nu \sigma} \slashed{x}
  \gamma^{\mu \rho} - 4 x^{\sigma} \gamma^{\mu \nu \rho} + 6 x^{\mu}
  \gamma^{\nu \rho \sigma} - 4 x^{\nu} \gamma^{\mu \sigma \rho}
  {+ 3 \gamma^{\mu} \slashed{x} \gamma^{\nu \rho \sigma}})\bigg] \, ,
    \end{aligned}
    \label{eq:pro}
\end{equation}
\end{widetext}
    where $m$ is the light quark mass, $i, j=1, 2, 3$ are color indices, and $T^a (a=1,...,8)$ is the Gell-Mann matrix. The related abbreviations are defined as 
\begin{equation}
    \begin{aligned}
  &\gamma^{\mu \nu \rho}  :=  \gamma^{\mu} \gamma^{\nu} \gamma^{\rho}\, , \quad
  g^{\{ \mu \rho } x^{ \sigma \}}  :=  g^{\mu \rho}
  x^{\sigma} + g^{\mu \sigma} x^{\rho} + g^{\rho \sigma} x^{\mu}\, ,\\
  &
  g^{\{ \mu \rho } \gamma^{ \sigma \}}  :=  g^{\mu \rho}
  \gamma^{\sigma} + g^{\mu \sigma} \gamma^{\rho} + g^{\rho \sigma}
  \gamma^{\mu}\, ,  \\
  &x^{\{ \mu }  x^{\rho} \gamma^{ \sigma \}}  :=  x^{\mu}
  x^{\rho} \gamma^{\sigma} + x^{\mu} x^{\sigma} \gamma^{\rho} + x^{\rho}
  x^{\sigma} \gamma^{\mu}\, , \\
  &G_{\mu \nu ; \rho}^a  := \tilde{D}_\rho^{ab} G^b_{\mu \nu}, G_{\mu \nu ; \rho \sigma}^a  := \tilde{D}_\sigma^{a b} \tilde{D}_\rho^{b c}  G^{c}_{\mu \nu}\, ,
    \end{aligned}
\end{equation}
in which $G_{\mu \nu}$ is the gluon filed strength, $\tilde{D}_\mu^{a b}=\delta^{a b}\partial_\mu-g f^{a b c} A^c_\mu$ is the covariant derivative operator in the adjoint representation, and $A^c_\mu$ is the external gauge field. These propagators can guarantee the infrared (IR) safety for the calculations involving the tri-gluon condensate $\langle g^3 f G^3 \rangle$~\cite{Zhang:2025fuz}.

\begin{figure*}
    \includegraphics[width=1.5\columnwidth]{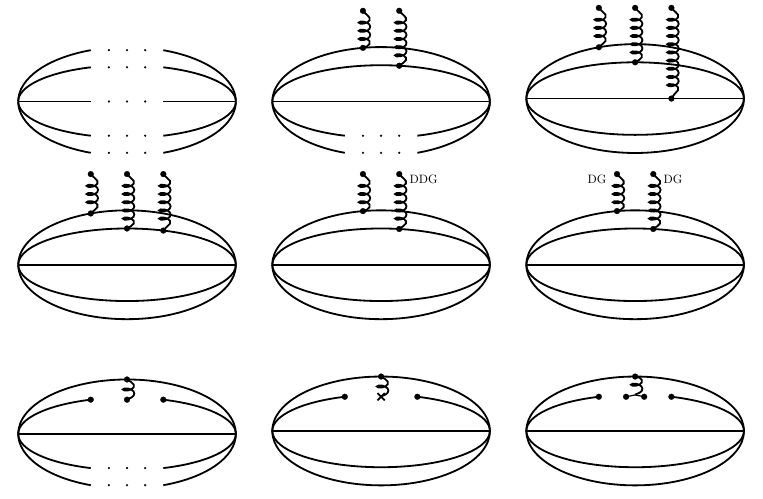}% Here is how to import EPS art
  \caption{The Feynman diagrams involved in our calculations for the $\Xi(1530)\bar{K}$ molecular pentaquark state. A quark line with dots contains terms proportional to the color factor $\delta^{ij}$ in the propagator $S^{ij}(x)$; consequently, such diagrams effectively subsume multiple Feynman diagrams.}
	\label{fig:massFeynman}
\end{figure*}

We preserve the calculations of spectral functions $\rho_{\slashed{p}}^{\text{OPE}}(s)$ and $\rho_{I}^{\text{OPE}}(s)$ up to dimension 13, and the corresponding Feynman diagrams are shown in Fig.~\ref{fig:massFeynman}. The results of $\rho_{\slashed{p}}^{\text{OPE}}(s)$ and $\rho_{I}^{\text{OPE}}(s)$ are too lengthy to present here, so we list them in Appendix~\ref{ope for 2012}. 

Due to the principle of quark-hadron duality, the correlation function computed at the quark-gluon level is expected to be equivalent to the one at the hadron level. Then the Borel transform is performed to suppress the contributions from high excited states. The hadron masses and coupling constants are obtained as
\begin{equation}
  \begin{aligned}&m_{\mp}^{2}(s_{0},M_{B}^2)\\&=\:\frac{\int_{s_{<}}^{s_{0}}[\sqrt{s}\rho_{\slashed{p}}^{\text{OPE}}(s)\pm\rho_{I}^{\text{OPE}}(s)]s~\mathrm{e}^{-s/M_{B}^{2}}\mathrm{d}s}{\int_{s_{<}}^{s_{0}}[\sqrt{s}\rho_{\slashed{p}}^{\text{OPE}}(s)\pm\rho_{I}^{\text{OPE}}(s)]~\mathrm{e}^{-s/M_{B}^{2}}\mathrm{d}s}
  \end{aligned}
  \label{mass}
\end{equation}
and
\begin{equation}
  \begin{aligned}&f_{\mp}^{2}(s_{0},M_{B})\\&=\:\frac{\int_{s_<}^{s_{0}}[\sqrt{s}\rho_{\slashed{p}}^{\text{OPE}}(s)\pm\rho_{I}^{\text{OPE}}(s)]\mathrm{e}^{-s/M_{B}^{2}}\mathrm{d}s}{2m_{\mp}}\times \mathrm{e}^{m_{\mp}^{2}/M_{B}^{2}}\, ,\end{aligned}
  \label{coupleconstant}
\end{equation}
which are functions of the Borel parameter $M_B$ and continuum threshold $s_0$. 

In the following analysis, we employ the same two-point QCD sum rule approach to determine the ground-state $\Xi$ baryon mass and its coupling constant $f_{\Xi}$ appearing in Eq.~\eqref{coplekandxi}.

\subsection{Three-point correlation function}
In QCD sum rules, the three-point correlation function is used to study the strong decay properties of hadrons. It is defined as
\begin{gather}
    \nonumber
    \Pi_{\mu}(p_1^2,p_2^2,p^2)\equiv \int \mathrm{d}^d x \mathrm{d}^d y ~\mathrm{e}^{\mathrm{i}p_1\cdot x}\mathrm{e}^{\mathrm{i}p_2\cdot y}\Gamma(x,y)\, ,\\
    \Gamma(x,y)\equiv\left\langle0|T[J_{\Xi}(x)J_{\bar{K}}(y)\bar{J}_{\mu}(0)]|0\right\rangle\, ,
    \label{therefunction}
\end{gather}
where $J_{\Xi}$ denotes the interpolating current associated with the $\Xi$ baryon 
\begin{equation}
    J_{\Xi}(x) =\epsilon ^{abc}\left[ s^{aT}( x) C\gamma _{\mu } s^{b}( x)\right] \gamma ^{\mu } \gamma ^{5} q^{c}( x)\, .
    \label{currentxi}
\end{equation}

The currents $J_{\Xi}$ and $J_{\bar{K}}$ couple to the physical states $|\Xi;1/2^+\rangle$ and $|\bar{K};0^-\rangle$ respectively, through the following matrix elements

\begin{gather}
    \nonumber
    \langle0|J_{\Xi}|\Xi;1/2^+\rangle=f_{\Xi}u(p_1)\, ,\\
    \langle0|J_{\bar{K}}|\bar{K};0^-\rangle=\frac{f_{\bar{K}}m_{\bar{K}}}{m_s+m_d}\equiv \lambda_{\bar{K}}\, .
    \label{coplekandxi}
\end{gather}
The interaction among the $\Omega^*$, $\Xi$, and $\bar{K}$ hadrons is described by the effective Lagrangian~\cite{Pascalutsa:2002pi}
\begin{equation}
\label{L}
    \mathcal{L}_{K \Xi\Omega^*}=\mathrm{i}g_{\Omega^*\Xi \bar{K}}\overline{\Xi}\gamma'^{\mu\nu\lambda}\gamma^5\left(\partial_\mu\Omega^*_\nu\right)\partial_\lambda \bar{K}+h.c.\, ,
\end{equation}
where $\gamma'^{\mu\nu\lambda}=i\varepsilon^{\mu\nu\lambda\alpha}\gamma_\alpha\gamma_5$.

After inserting complete set of intermediate hadronic states, the phenomenological representation of the three-point correlation function can be written as
\begin{align}
  \nonumber
    \Pi&_{\mu}(p_1^2,p_2^2,p^2)= \\
    \nonumber
    & \frac{\lambda_{\bar{K}}f_{\Xi}u(p_1)\langle  \bar{K}(p_2)\Xi(p_1) | \Omega^*(p)\rangle f_{-}\bar{u}_\mu(p)}{(p^2-m_{\Omega^*}^{2}+\mathrm{i}\varepsilon)(p_1^2-m_\Xi^2+\mathrm{i}\varepsilon)(p_2^2-m_{\bar{K}}^2+\mathrm{i}\varepsilon)} +...\, ,\\
    \label{threefunciotnphe}
\end{align}
where the “...” contains the contributions of higher excited states, including those from opposite-parity states. After the Borel transformation, the ground-state contributions are expected to dominate the sum rules. Since both $\Omega^*(3/2^-)$ and $\Xi(1/2^+)$ are treated as ground states in the corresponding channels, the contaminations from opposite-parity states are therefore suppressed and are not expected to significantly affect the numerical results. The transition matrix element can be derived from the effective Lagrangian in Eq.~\eqref{L} as
\begin{equation}
    \langle  \bar{K}(p_2)\Xi(p_1) |{\Omega^*}(p)\rangle=\bar{u}(p_1)\mathrm{i}g_{{\Omega^*}\Xi \bar{K}}\gamma'^{\mu\nu\lambda}\gamma^5p_2^{\lambda}p^{\mu}u_{\nu}(p)\, .
    \label{tmatrix}
  \end{equation}
Substituting Eq.~\eqref{tmatrix} into Eq.~\eqref{threefunciotnphe}, we obtain
\begin{align}
  \nonumber
    \Pi&_{\mu}(p_1^2,p_2^2,p^2) \\
    \nonumber
    =& \frac{\mathrm{i}g_{{\Omega^*}\Xi \bar{K}}\lambda_{\bar{K}}f_{\Xi}f_{-}m_\Xi p_2 ^2 \slashed{p}_1\gamma^\mu\gamma^5}{3(p^2-m_{\Omega^*}^2+\mathrm{i}\epsilon)(p_1^2-m_\Xi^2+\mathrm{i}\varepsilon)(p_2^2-m_{\bar{K}}^2+\mathrm{i}\varepsilon)}\\
    \nonumber+...\,  .\\
    \label{therefunctionre}
\end{align}
In view of the complexity of the full expression, we keep only the Lorentz structure $\slashed{p}_1 \gamma^{\mu}\gamma^{5}$ in Eq.~\eqref{therefunctionre} for the convenience of subsequent analysis.

This result can be equivalently rewritten as
\begin{align}
\nonumber
    \Pi&_{\mu}(p_1^2,p_2^2,p^2) = \\
  \nonumber
    & (1+\frac{m^2_{\bar{K}}}{p_2^2-m_{\bar{K}}^2+\mathrm{i}\varepsilon})\frac{\frac{1}{3}m_\Xi \mathrm{i}g_{{\Omega^*}\Xi \bar{K}}\lambda_{\bar{K}}f_{\Xi}f_{-}}{(p^2-m_{\Omega^*}^2+\mathrm{i}\varepsilon)(p_1^2-m_\Xi^2+\mathrm{i}\varepsilon)}\\
    \nonumber
    &+...\,  .\\
    \label{therefunctionre1}
\end{align}

\begin{figure*}
    \includegraphics[width=2\columnwidth]{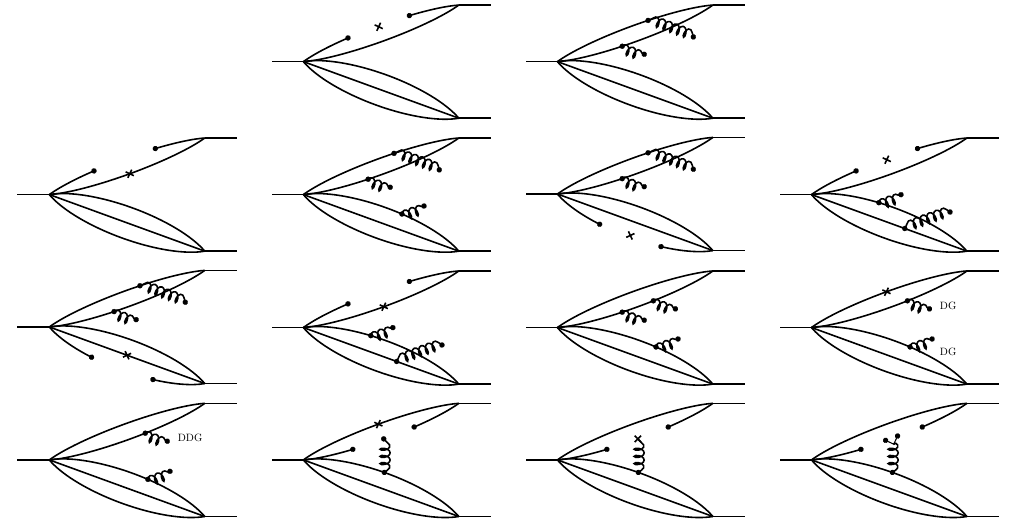}% Here is how to import EPS art
  \caption{The Feynman diagrams which contribute to the theoretical side of the three-point sum rules in the relevant  structure for the two-body strong decays of $\Xi(1530)\bar{K}$ molecular pentaquark state. }
	\label{fig:decayFeynman}
\end{figure*}

Since $p = p_1 + p_2$, we consider the soft-kaon limit $p_2 \to 0$, in which case $p^2 \approx p_1^2$. In this limit, the terms proportional to $1/p_2^2$ become dominant in the OPE series. Therefore, to properly construct the sum rules, we isolate the contributions with the Lorentz structure $\slashed{p}_1 \gamma^\mu \gamma^5$ proportional to $1/p_2^2$. This approximation has been examined in Ref.~\cite{Eidemuller:2005jm}, where the results obtained in the soft-meson limit were found to be consistent with those treating the meson momentum as a small but finite quantity. This procedure determines the coupling constant $g_{{\Omega^*}{\Xi}{\bar{K}}}(Q^2)$ (where $Q^2 = -p_2^2$) in a region away from the physical pole at $Q^2 = -m_{\bar{K}}^2$. The on-shell coupling is then obtained by extrapolating from the valid QCD sum rule window. By equating the hadronic and OPE representations, we have
\begin{align}
\nonumber
    & \frac{m^2_{\bar{K}}}{(p_2^2-m_{\bar{K}}^2+\mathrm{i}\varepsilon)}\frac{\frac{1}{3}m_\Xi \mathrm{i}g_{{\Omega^*}{\Xi}{\bar{K}}}\lambda_{\bar{K}}f_{\Xi}f_{-}}{(p^2-m_{\Omega^*}^2+\mathrm{i}\varepsilon)(p^2-m_\Xi^2+\mathrm{i}\varepsilon)}+...\\ 
    \nonumber 
    &=\frac{1}{p^2_2}\left[\int_{0}^{s_0}\mathrm{d}s\frac{\rho(s)}{s-p^2}+\int_{s_0}^{\infty}\mathrm{d}s\frac{\rho(s)}{s-p^2}\right]\, .\\
    \label{therefunctionre2}
\end{align}

According to quark-hadron duality, the contributions from higher resonances and the continuum on the phenomenological side are assumed to be dual to the integral above the continuum threshold $s_0$ on the OPE side. After performing the Borel transform, the sum rules are obtained as
\begin{align}
  \nonumber
    &\frac{1}{3}m_\Xi m^2_{\bar{K}} \mathrm{i}g_{{\Omega^*}{\Xi}{\bar{K}}}\lambda_{\bar{K}}f_{\Xi}f_{-} \frac{Q^2+m^2_{\bar{K}}}{Q^2}\frac{\mathrm{e}^{-m_{{\Omega^*}}^2/M^2_B}-\mathrm{e}^{-m_{\Xi}^2/M^2_B}}{m_{{\Omega^*}}^2-m_{\Xi}^2}\\ 
    \nonumber 
    &=\int_{0}^{s_0}\mathrm{d}s\rho(s)\mathrm{e}^{-s/M^2_B}\, .\\
    \label{decay}
\end{align}

To extract the value of the coupling constant at the physical point $Q^2 = -m_{\bar{K}}^2$, we extrapolate $g_{{\Omega^*}{\Xi}{\bar{K}}}(Q^2)$ from the working region of QCD sum rules by using an exponential model~\cite{Chen:2015fsa,Dias:2013xfa,Lian:2023cgs}
\begin{equation}
    g_{{\Omega^*}{\Xi}{\bar{K}}}(Q^2)=g_1\mathrm{e}^{-g_2Q^2}\, ,
    \label{region}
\end{equation}
where the parameters $g_1$ and $g_2$ are determined by fitting the numerical results obtained from the sum rules.

The decay width of the process $\Omega^*\rightarrow{\Xi}{\bar{K}}$ can be derived from Eq.~\eqref{L} as
\begin{align}
\nonumber
\Gamma(\Omega^*\rightarrow{\Xi}{\bar{K}})
    =&\frac{\sqrt{\lambda(m_{{\Omega^*}}^2,m_{\Xi}^2,m_{\bar{K}}^2)}}{192\pi m_{{\Omega^*}}^3}g^2_{{\Omega^*}{\Xi}{\bar{K}}}\\
    \nonumber
    \times&(m_{{\Omega^*}}^2-2m_{{\Omega^*}}m_{\Xi}-m_{\bar{K}}^2+m_{\Xi}^2)^2\\
    \nonumber
    \times&(m_{{\Omega^*}}^2+2m_{{\Omega^*}}m_{\bar{K}}-m_{\bar{K}}^2+m_{\Xi}^2)\, ,\\
    \label{decatre}
\end{align}
where $\lambda$ is the well-known K\"all\'en function
\begin{equation}
    \lambda(a,b,c)=a^2+b^2+c^2-2ab-2ac-2bc\, .
\end{equation}

At the quark–gluon level, we evaluate the three-point correlation function using the same light-quark propagators given in Eq.~\eqref{eq:pro}. We preserve the spectral function $\rho^{\text{OPE}}(s)$ up to dimension 10 and the corresponding Feynman diagrams are shown in Fig.~\ref{fig:decayFeynman}. The result of $\rho(s)$ is obtained as
\begin{equation}
\label{rhodecay}
\begin{aligned}
\rho^{\text{OPE}}(s)=&\frac{5\langle g^{2}G^{2}\rangle}{98304\pi^{6}}s^{2}+\frac{5m_{s}\langle\overline{s}s\rangle}{6144\pi^{4}}s^{2}-\frac{5m_{s}\langle\overline{q}q\rangle}{3072\pi^{4}}s^{2}\\
-&\frac{7  g^{2}\langle\overline{s}s\rangle^{2}}{41472\pi^{4}}s-\frac{7 g^{2}\langle\overline{q}q\rangle^{2}}{41472\pi^{4}}s-\frac{7  \langle g^3fG^{3}\rangle}{2^{15}\times3^3\times\pi^{6}}s\\
-&\frac{7  m_{s}\left\langle \bar{s}g\sigma Gs\right\rangle}{18432\pi^{4}}s+\frac{m_{s}\langle \bar{q}g\sigma Gq\rangle}{1536\pi^{4}}s\\
-&\frac{55m_{s}\langle\overline{s}s\rangle\langle g^{2}G^{2}\rangle}{12288\pi^{4}}+\frac{5m_{s}\langle\bar{q}q\rangle\langle g^{2}G^{2}\rangle}{6144\pi^{4}}\, .
\end{aligned}
\end{equation}

\section{Numerical analyses}
\label{sec:numerical}
\subsection{The mass of $\Xi(1530)\bar{K}$ molecular pentaquark}
For the numerical analyses, we adopt values of the QCD parameters at the renormalization scale $\mu = 1~\mathrm{GeV}$ with $\Lambda_{\rm QCD}=300~\mathrm{MeV}$, which are standard in QCD sum rule calculations~\cite{Ioffe:1981kw,Narison:1988ep,Jamin:2001zr,Jamin:1998ra,Dosch:1988vv,Khodjamirian:2011ub,ParticleDataGroup:2024cfk,Francis:2018jyb,Narison:2018nbv,Chung:1984gr}
\begin{equation}
  \begin{aligned}
  m_u&=m_d=0\, ,m_s=93.5\pm0.8~\mathrm{MeV}\, ,\\
  \left\langle\bar{q}q\right\rangle&=-(0.24\pm0.01)^3~\mathrm{GeV^3}\, ,\\
  \left\langle\bar{s}s\right\rangle&=(0.8\pm 0.1)\left\langle\bar{q}q\right\rangle\, ,\\
  \left\langle\bar{q}g\sigma Gq\right\rangle&=(0.8\pm 0.2)\left\langle\bar{q}q\right\rangle ~\mathrm{GeV^2}\, ,\\
  \left\langle g^2 G^2 \right\rangle&=(0.48\pm 0.14) ~\mathrm{GeV^4}\, ,\\
  \left\langle g^3fG^3 \right\rangle&=(8.2\pm 1)\left\langle \alpha_s G^2 \right\rangle~\mathrm{GeV^2}\, .
  \end{aligned}
  \label{values}
\end{equation}
Here the strange quark mass is taken at the renormalization scale  $\mu = 2~\mathrm{GeV}$. The renormalization group equation is used to evolve $m_s$ to the working scale $\mu = 1~\mathrm{GeV}$.

\begin{figure}[h]
  \includegraphics[width=\columnwidth]{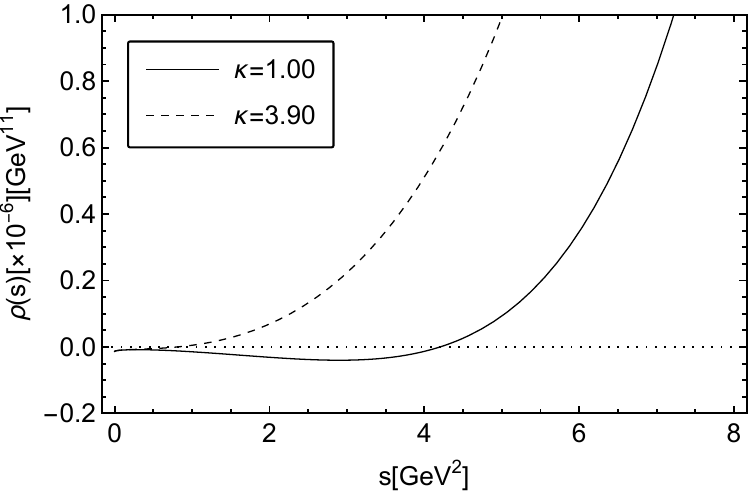}% Here is how to import EPS art
  \caption{\label{fig:densities} Behaviors of the spectral densities for
different factorization assumption. The solid lines represent the spectral densities for
$\kappa=1$, whereas the dashed lines are the corresponding densities considering $\kappa=3.90$.}
\end{figure}

We begin by examining the behavior of the spectral function before proceeding to the mass sum rule analysis. As illustrated in Fig.~\ref{fig:densities}, the spectral density suffers from poor positivity, becoming negative over a relatively wide region, $1~\mathrm{GeV}^2 \leq s \leq 4~\mathrm{GeV}^2$. To remedy this unphysical behavior, we relax the factorization assumption by introducing a parameter $\kappa$ via $\left\langle\bar{q}\bar{q}qq\right\rangle=\kappa\left\langle\bar{q}q\right\rangle^2$~\cite{Narison:2009vy,Narison:1995jr,Chung:1984gr}. With $\kappa=3.90$, the spectral density exhibits sufficiently good behavior, which will be adopted in the following analysis.  The associated uncertainty is estimated by varying $\kappa$ within $\kappa = 3.90 \pm 0.30$, and is included in the final error estimates.

\begin{figure}[h]
  \includegraphics[width=\columnwidth]{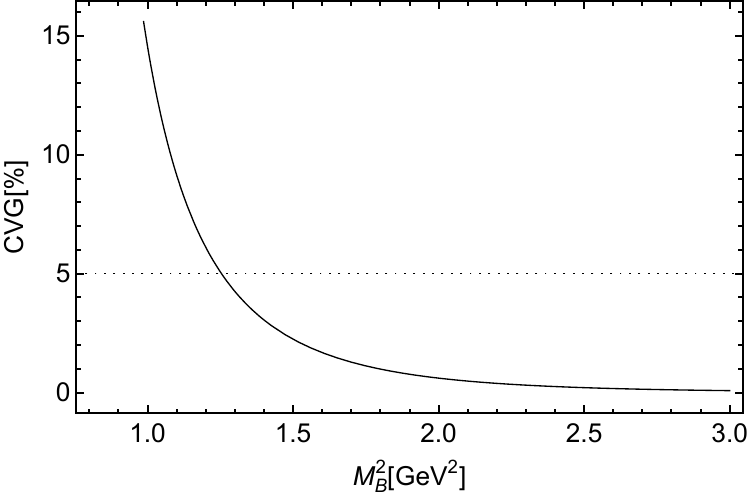}% Here is how to import EPS art
  \caption{\label{fig:OPE2012} OPE convergence for the $\Xi(1530)\bar{K}$ molecular pentaquark state considering $\kappa=3.90$.}
\end{figure}
\begin{figure}[h]
  \includegraphics[width=\columnwidth]{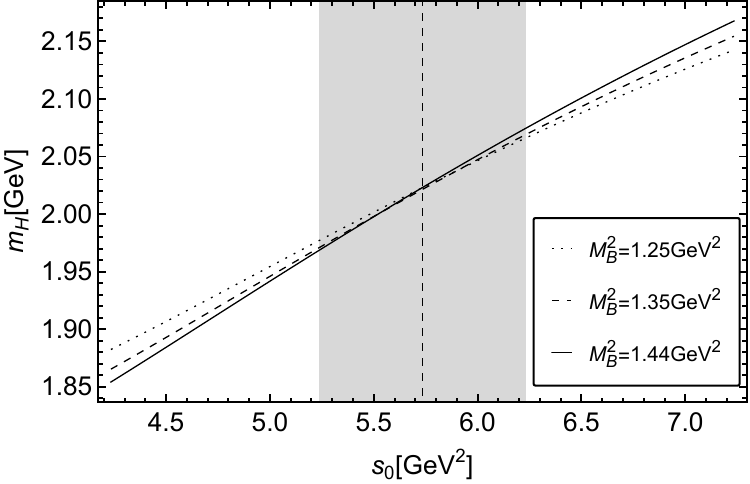}% Here is how to import EPS art
  \caption{\label{fig:masss02012} Variation of hadron mass with $s_0$ corresponding to the $\Xi(1530)\bar{K}$ molecular pentaquark state considering $\kappa=3.90$.}
\end{figure}
\begin{figure}[h]
  \includegraphics[width=\columnwidth]{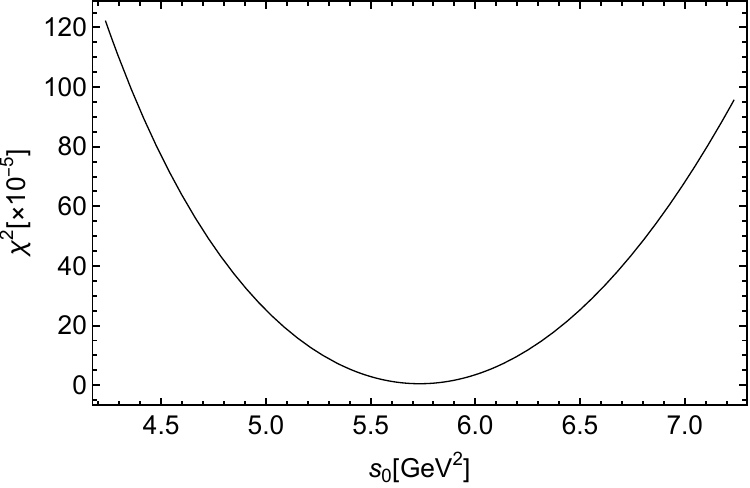}% Here is how to import EPS art
  \caption{\label{fig:chi2012} Variation of $\chi^2$ with $s_0$ corresponding to the $\Xi(1530)\bar{K}$ molecular pentaquark state considering $\kappa=3.90$.}
\end{figure}
\begin{figure}[h]
  \includegraphics[width=\columnwidth]{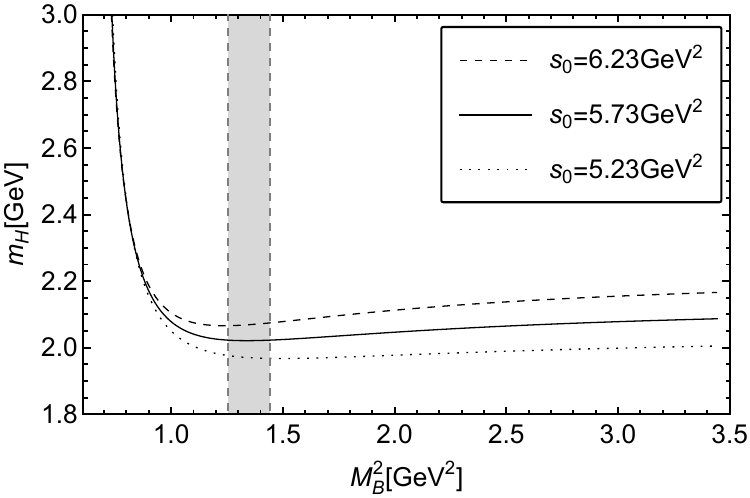}% Here is how to import EPS art
  \caption{\label{fig:massmb2012}Variation of hadron mass with $M^2_B$ corresponding to the $\Xi(1530)\bar{K}$ molecular pentaquark state considering $\kappa=3.90$.}
\end{figure}

In Eq.~\eqref{mass}, the extracted hadron mass depends on two free parameters: the Borel parameter $M_B^2$ and continuum threshold $s_0$. Their working regions are determined by imposing the following standard criteria:
(a) good convergence of the operator product expansion, 
(b) a sufficiently large pole contribution, and
(c) good stability of the mass sum rules.

The convergence of the OPE is quantified by requiring that the contributions from high-dimensional operators be sufficiently suppressed
\begin{equation}
\mathrm{CVG}\equiv\left|\frac{\Pi_-^{12+13}(\infty,M_B^2)}{\Pi_-(\infty,M_B^2)}\right|\leq5\%\, ,
\label{CVG2012}
\end{equation}
where $\Pi_-$ denotes the full OPE series of correlation function, while $\Pi_-^{12+13}$ represents the combined contributions from the dimension-12 and 13 condensates. As shown in Fig.~\ref{fig:OPE2012}, this requirement leads to a lower bound on the Borel parameter $M_B^2 \geq 1.25~\mathrm{GeV}^2$. 

To determine the optimal value of the continuum threshold $s_0$, we plot the $m_H$–$s_0$ curves for different values of $M_B^2$ in Fig.~\ref{fig:masss02012}. The optimal $s_0$ is chosen such that the dependence of $m_H$ on $M_B^2$ is minimized in its vicinity. To quantify this criterion, we define the function $\chi^2$ as~\cite{Xu:2025oqn,Li:2025hsp,Li:2025dkw,Chen:2026ybf}
\begin{equation}
\chi^2(s_0)=\sum_{i=1}^{N}\left[\frac{m_H(s_0,M^2_{B,i})}{\bar{m}_H(s_0)}-1\right]^2\, ,
\label{eq:chi}
\end{equation}
where $\bar{m}_H(s_0)$ denotes the average value of the data points
\begin{equation}
\bar{m}_H(s_0)=\sum_{i=1}^{N}\frac{m_H(s_0,M^2_{B,i})}{N}\, .
\label{eq:mave}
\end{equation}
The $\chi^2$ as a function of $s_0$ is shown in Fig.~\ref{fig:chi2012}, from which the optimal value of the continuum threshold is determined from the location of the minimum as $s_0=5.73$ GeV$^2$.

In addition, for a sufficient pole contribution we require
\begin{equation}
  \mathrm{PC}\equiv\left|\frac{\Pi_-(s_0,M_B^2)}{\Pi_-(\infty,M_B^2)}\right|\geq50\%\, .
  \label{pc2012}
\end{equation}
Applying the above constraints yields an acceptable Borel window of $1.25~\mathrm{GeV}^2 \leq M_B^2 \leq 1.44~\mathrm{GeV}^2$. The corresponding Borel curves are displayed in Fig.~\ref{fig:massmb2012}, where the extracted mass exhibits good stability against variations of $M_B^2$ within this window. As a result, we obtain the mass and coupling constant of the $\Xi(1530)\bar{K}$ molecular state as
\begin{gather}
\label{mass2012}
    m_{\Xi(1530)K}=2.02\pm0.12~\mathrm{GeV}\, ,\\   
    f_{-}=(4.60\pm1.17)\times10^{-4}~\mathrm{GeV}^6\, , 
  \label{decayconstant2012}
\end{gather}
where the errors are mainly from the uncertainties of the various condensates in Eq.~\eqref{values}, as well as the phenomenological parameter $\kappa$. The extracted mass of the $\Xi(1530)\bar{K}$ molecular petaquark state is consistent with the mass of $\Omega(2012)$ collected in PDG~\cite{ParticleDataGroup:2024cfk}.

\subsection{The coupling constant of $\Xi$}
\begin{figure}[h]
  \includegraphics[width=\columnwidth]{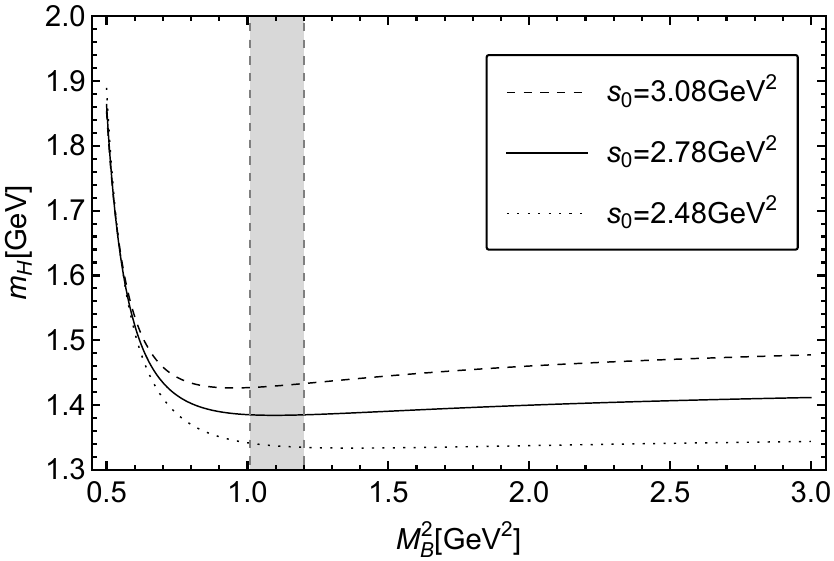}% Here is how to import EPS art
  \caption{\label{fig:massmbxi}Variation of hadron mass with $M^2_B$ corresponding to the $\Xi$ state.}
\end{figure}

To calculate the coupling constant of $\Xi$, we perform the mass sum rule analysis for the current in Eq.~\eqref{currentxi} and adopt the same values of the QCD parameters in Eq.~\eqref{values}. Under the following constraints
\begin{gather}
\nonumber
    \mathrm{CVG}\equiv\left|\frac{\Pi_-^{6+7+8}(\infty,M_B^2)}{\Pi_-(\infty,M_B^2)}\right|\leq10\%\, ,\\
    \mathrm{PC}\equiv\left|\frac{\Pi_-(s_0,M_B^2)}{\Pi_-(\infty,M_B^2)}\right|\geq40\%\, ,
    \label{cvgandpcxi}
\end{gather}
the acceptable Borel window is determined to be
$1.01~\mathrm{GeV}^2 \leq M_B^2 \leq 1.20~\mathrm{GeV}^2$.
The corresponding Borel curves are shown in Fig.~\ref{fig:massmbxi}. As a result, we obtain the mass and coupling constant of $\Xi$ as

\begin{gather}
\label{massxi}
m_{\Xi}=1.38\pm0.06~\mathrm{GeV}\, ,\\
  f_{\Xi}=(3.21\pm0.39)\times10^{-2}~\mathrm{GeV^3}\, ,
  \label{decayconstantxi}
\end{gather}
in which the obtained mass agrees well with the mass of $\Xi$ baryon with $J^P=1/2^+$ in PDG~\cite{ParticleDataGroup:2024cfk}.

\subsection{The strong decays of $\Omega(2012)\rightarrow \Xi^0 K^-$ and $\Xi^-\bar{K}^0$}
To perform the three-point sum rule analyses, we adopt the input parameters listed in Eq.~\eqref{values} and the same factorization parameter $\kappa=3.90$. For the  $\Xi$ and $\bar{K}$ doublets and $\Omega(2012)$, we use the hadron parameter values from PDG~\cite{ParticleDataGroup:2024cfk}: 
\begin{equation}
  \begin{aligned}
m_{\Omega(2012)}&=2012.4\pm0.9~\mathrm{MeV}\, ,\\
m_{\Xi^-}&=1321.71\pm0.07~\mathrm{MeV}\, , \\
m_{\Xi^0}&=1314.86\pm0.20~\mathrm{MeV}\, , \\
m_{\bar{K}^0}&=497.611\pm0.013~\mathrm{MeV}\, , \\
m_{K^-}&=493.677\pm0.015~\mathrm{MeV}\, , \\
f_{\bar{K}}&=155.7~\mathrm{MeV}\, .
  \end{aligned}
  \label{hadronparametervalues}
\end{equation}

\begin{figure}[h]
  \includegraphics[width=\columnwidth]{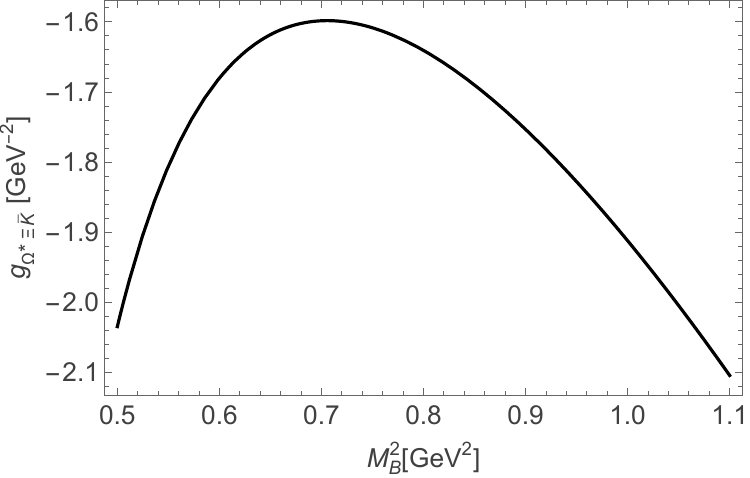}
  \caption{\label{fig:g} Variation of the coupling constant $g_{{\Omega^*}\Xi^- \bar{K}^0}(Q^2 = m^2_{\Xi^-})$ with the Borel parameter $M_B^2$ for $s_0=5.73$ GeV$^2$.}
\end{figure}

As shown in Eq.~\eqref{decay}, the coupling constant extracted from the sum rule depends on the $s_0$ and $M_B^2$ in the limit $Q^2 \to 0$.
Adopting the same value of $s_0=5.73$ GeV$^2$ in the mass sum rule, 
we present the dependence of the coupling constant on $M^2_B$ at the Euclidean  momentum point $Q^2 = m^2_{\Xi^-}$ in Fig.~\ref{fig:g}. We find that the sum rules yield a stable platform for the coupling constant at $M_B^2= 0.71~\mathrm{GeV^2}$, around which it has minimal parameter dependence.
\begin{figure}[h]
  \includegraphics[width=\columnwidth]{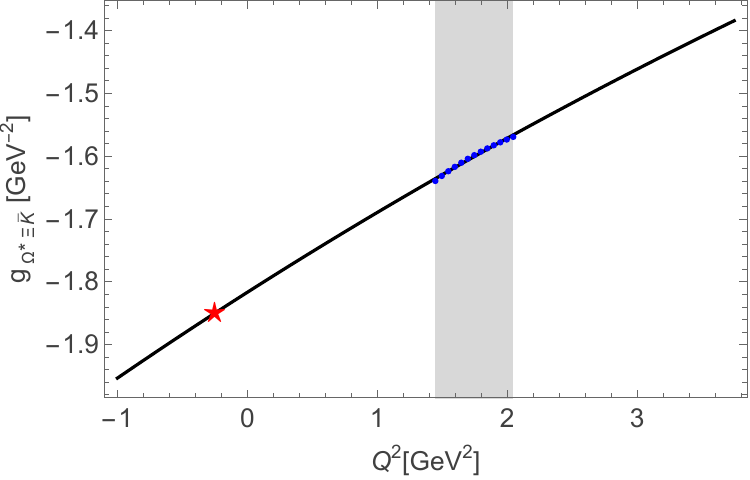}
  \caption{\label{fig:gQ} Variation of the coupling constant $g_{{\Omega^*}\Xi^- \bar{K}^0}$ with $Q^2$. The dotted points represent the QCD sum rules results for $g_{{\Omega^*}\Xi^- \bar{K}^0}$ with $s_0=5.73~\mathrm{GeV^2}$ and $M_B^2=0.71~\mathrm{GeV^2}$. The solid line is the fit of the QCD sum rules result through model Eq.~\eqref{region} and the extrapolation to the physical pole $Q^2=-m_{\bar{K}^0}^2$.}
\end{figure}

To extract the value of the coupling constant at the physical point $Q^2 = -m_{\bar{K}^0}^2$, we extrapolate $g_{{\Omega^*}\Xi^- \bar{K}^0}(Q^2)$ from the QCD sum rules working region using the exponential model given in Eq.~\eqref{region}. As shown in Fig.~\ref{fig:gQ}, the QCD sum rules results obtained with $s_0=5.73~\mathrm{GeV}^2$ and $M_B^2=0.71~\mathrm{GeV}^2$ are well described by this model, yielding the fitted parameters $g_1=-1.82~\mathrm{GeV}^{-2}$ and $g_2=0.07~\mathrm{GeV}^{-2}$. Using these parameters, the coupling constant at $Q^2=-m_{\bar{K}^0}^2$ is obtained as
\begin{equation}
    g_{{\Omega^*}\Xi^- \bar{K}^0}=-1.85^{+0.87}_{-0.69}~\mathrm{GeV}^{-2}\, ,
    \label{g2012}
\end{equation}
where the errors are mainly from the uncertainties of the coupling constant in Eqs.~\eqref{decayconstant2012} and \eqref{decayconstantxi}, the various condensates in Eq.~\eqref{values}, and the hadron parameter values in Eq.~\eqref{hadronparametervalues}, as well as the phenomenological parameter $\kappa$.

According to Eq.~\eqref{decatre}, the partial decay width of the process $\Omega(2012)\rightarrow \Xi^-\bar{K}^0$ can be calculated as
\begin{equation}
    \Gamma({\Omega^*}\rightarrow \Xi^-\bar{K}^0)=0.44^{+0.51}_{-0.27}~\mathrm{MeV}\, .
    \label{decay2012}
\end{equation}

% For the $\Omega(2012)\rightarrow \Xi^0K^-$ decay, the only difference comes from the isospin mass-splitting of the $\Xi$ and $K$ doublets in the final states.

For the $\Omega(2012)\rightarrow \Xi^0K^-$ decay, the dominant difference arises from the isospin mass splitting of the $\Xi$ and $\bar{K}$ doublets in the final states. Possible isospin-breaking effects in the interpolating current and the OPE, such as those induced by $m_u \neq m_d$, are neglected due to their small numerical impact.  Following the same procedure, we obtain the fitted parameters $g_1 = 1.85~\mathrm{GeV}^{-2}$ and $g_2 = 0.07~\mathrm{GeV}^{-2}$ in the extrapolation model. With these values, the coupling constant at $Q^2 = -m_{K^-}^2$ and the corresponding partial decay width are determined to be
\begin{gather}
    g_{{\Omega^*}\Xi^0 K^-}=1.88^{+0.88}_{-0.70}~\mathrm{GeV}^{-2}\, ,\\
    \Gamma({\Omega^*}\rightarrow \Xi^0K^-)=0.52^{+0.60}_{-0.31}~\mathrm{MeV}\, .
    \label{decay2012v2}
\end{gather}

As a result, the total two-body decay width for the $\Xi^\ast \bar{K}$ molecular pentaquark state are
\begin{align}
\Gamma({\Omega^*}\rightarrow\Xi \bar{K})=0.96^{+0.79}_{-0.41}~\mathrm{MeV}\, .\label{totalwidty}
\end{align}
This value agrees with the theoretical predictions from the effective Lagrangian approach in Refs.~\cite{Huang:2018wth,Lin:2018nqd,Lin:2019tex}. Given the neglect of three-body decays, it is also roughly consistent with the results reported by the ALICE ~\cite{ALICE:2025atb}.
The branching fraction ratio of the two-body decays is obtained as
\begin{align}
\nonumber
\mathcal{R}^{\Xi^-\bar{K}^0}_{\Xi^0K^-}\equiv &\frac{\mathcal{B}[\Omega(2012)\rightarrow \Xi^-\bar{K}^0]}{\mathcal{B}[\Omega(2012)\rightarrow \Xi^0K^-]} \\ =&\frac{\Gamma({\Omega^*}\rightarrow \Xi^-\bar{K}^0)}{\Gamma({\Omega^*}\rightarrow \Xi^0K^-)}\approx 0.85\, ,
\end{align}
which is in good agreement with the experimental data collected in PDG~\cite{ParticleDataGroup:2024cfk}.

\section{CONCLUSION AND DISCUSSION}
\label{sec:summary}
In this work, we have investigated the recently observed $\Omega(2012)$ as a $\Xi(1530)\bar{K}$ molecular pentaquark state within the framework of QCD sum rules. We construct the unique molecular interpolating current with $I(J^P)=0(\frac{3}{2}^-)$ and calculate the correlation functions by employing the parity-projected sum rules to separate the positive- and negative-parity contributions. We calculate the two-point correlation function to investigate the hadron mass and the three-point correlation function to study the two-body strong decay properties of the $\Xi(1530)\bar{K}$ pentaquark state.

Our QCD sum rule analyses yield a mass of $2.02\pm0.11~\mathrm{GeV}$ for the $\Xi(1530)\bar{K}$ molecular pentaquark state, a total two-body decay width of $\Gamma({\Omega^*} \rightarrow \Xi \bar{K}) = 0.96^{+0.79}_{-0.41}~\mathrm{MeV}$. The branching fraction ratio of the two-body decay channels is obtained as $\mathcal{R}^{\Xi^-\bar{K}^0}_{\Xi^0K^-} = 0.85$. These results are in agreement with the experimental data for $\Omega(2012)$ within uncertainties, supporting its interpretation as a $\Xi(1530)\bar{K}$ molecular pentaquark state. These calculations are useful for providing further insight into the internal structure of $\Omega(2012)$ in the future.

\section*{Acknowledgments}
This work is supported by the National Natural Science Foundation of China under Grant No. 12575153. 

\appendix
\section{The spectral densities for the negative-parity $\Xi(1530)\bar{K}$ molecular state  }
\label{ope for 2012}
The explicit expressions for the spectral densities $\rho_{\slashed{p}}^{\text{OPE}}(s)$ and $\rho_{I}^{\text{OPE}}(s)$ are presented in Eqs.~\eqref{rhop} and \eqref{rhoi}, respectively.
\textcolor{red}{}
\begin{widetext}
\begin{equation}
  \begin{aligned}
    \rho_{\slashed{p}}^{\text{OPE}}(s)=&\frac{  s^5}{15728640 \pi ^8}+\frac{31 \langle\bar{s}s\rangle   m_s s^3}{737280 \pi ^6}-\frac{13 \langle\bar{q}q\rangle   m_s s^3}{92160 \pi ^6}+\frac{61 \langle g^{2}G^{2}\rangle   s^3}{212336640 \pi ^8}+\frac{251   \langle\bar{q}g\sigma Gq\rangle m_s s^2}{368640 \pi ^6}\\
    +&\frac{  \langle\bar{s}g\sigma Gs\rangle m_s s^2}{6144 \pi ^6}-\frac{ g^{2} \langle\bar{q}q\rangle^2   s^2}{41472 \pi ^6}-\frac{83  g^{2} \langle\bar{s}s\rangle^2   s^2}{622080 \pi ^6}-\frac{\langle\bar{q}q\rangle^2   s^2}{4608 \pi ^4}+\frac{7 \langle\bar{s}s\rangle^2   s^2}{11520 \pi ^4}+\frac{\langle\bar{q}q\rangle \langle\bar{s}s\rangle   s^2}{360 \pi ^4}\\
    -&\frac{23 \langle g^3G^{3}\rangle   s^2}{5308416 \pi ^8}-\frac{35 \langle g^{2}G^{2}\rangle \langle\bar{q}q\rangle   m_s s}{110592 \pi ^6}-\frac{329 \langle g^{2}G^{2}\rangle  \langle\bar{s}s\rangle   m_s s}{589824 \pi ^6}+\frac{\langle\bar{q}q\rangle   \langle\bar{q}g\sigma Gq\rangle s}{4608 \pi ^4}\\
    -&\frac{37 \langle\bar{q}q\rangle   \langle\bar{s}g\sigma Gs\rangle s}{4608 \pi ^4}-\frac{49 \langle\bar{s}s\rangle   \langle\bar{q}g\sigma Gq\rangle s}{6912 \pi ^4}-\frac{143 \langle\bar{s}s\rangle   \langle\bar{s}g\sigma Gs\rangle s}{27648 \pi ^4}-\frac{\langle\bar{q}q\rangle   \langle\bar{s}g\sigma Gs\rangle m_s^2}{864 \pi ^4}\\
    -&\frac{91 \langle\bar{s}s\rangle   \langle\bar{s}g\sigma Gs\rangle m_s^2}{6912 \pi ^4}+\frac{2405 \langle g^{2}G^{2}\rangle   \langle\bar{s}g\sigma Gs\rangle m_s}{5308416 \pi ^6}+\frac{985 \langle g^{2}G^{2}\rangle   \langle\bar{q}g\sigma Gq\rangle m_s}{5308416 \pi ^6}+\frac{\langle\bar{s}s\rangle^3   m_s}{96 \pi ^2}\\
    -&\frac{13 \langle\bar{q}q\rangle^2 \langle\bar{s}s\rangle   m_s}{96 \pi ^2}-\frac{5  g^{2} \langle\bar{q}q\rangle^2 \langle\bar{s}s\rangle   m_s}{1152 \pi ^4}+\frac{5 g^{2} \langle\bar{q}q\rangle \langle\bar{s}s\rangle^2   m_s}{1296 \pi ^4}-\frac{11 g^{2} \langle\bar{s}s\rangle^3   m_s}{1944 \pi ^4}\\
    -&\frac{ g^{2} \langle\bar{q}q\rangle^3   m_s}{2592 \pi ^4}+\frac{11 \langle g^3G^{3}\rangle \langle\bar{q}q\rangle   m_s}{24576 \pi ^6}+\frac{113 \langle g^3G^{3}\rangle \langle\bar{s}s\rangle   m_s}{147456 \pi ^6}-\frac{11 g^{2} \langle g^{2}G^{2}\rangle \langle\bar{q}q\rangle^2  }{2239488 \pi ^6}\\
    -&\frac{\langle g^{2}G^{2}\rangle \langle\bar{q}q\rangle^2  }{6912 \pi ^4}+\frac{11 \langle g^{2}G^{2}\rangle \langle\bar{s}s\rangle^2  }{9216 \pi ^4}+\frac{41 \langle g^{2}G^{2}\rangle \langle\bar{s}s\rangle^2  }{8957952 \pi ^6}\\
    +&\frac{11 \langle g^{2}G^{2}\rangle \langle\bar{q}q\rangle \langle\bar{s}s\rangle  }{13824 \pi ^4}+\frac{193   \langle\bar{q}g\sigma Gq\rangle \langle\bar{s}g\sigma Gs\rangle}{27648 \pi ^4}+\frac{5   \langle\bar{q}g\sigma Gq\rangle^2}{27648 \pi ^4}+\frac{3   \langle\bar{s}g\sigma Gs\rangle^2}{1024 \pi ^4}\\
    -&\frac{5 {\langle g^{2}G^{2}\rangle} {\langle\bar{q}q\rangle} {\langle\bar{q}g\sigma Gq\rangle}  \delta (s)}{27648 \pi ^3}-\frac{5 {g}^4 {\langle\bar{q}q\rangle}^2   {\langle\bar{s}s\rangle}^2\delta (s)}{139968 \pi ^3}-\frac{5 {g}^4 {\langle\bar{q}q\rangle}^4  \delta (s)}{839808 \pi ^3}-\frac{11 {g}^2 {\langle\bar{q}g\sigma Gq\rangle}   {\langle\bar{s}s\rangle}^2 {m_s}\delta (s)}{20736 \pi ^3}\\
    -&\frac{11 {g}^2 {\langle\bar{q}q\rangle}^2 {\langle\bar{q}g\sigma Gq\rangle}   {m_s}\delta (s)}{31104 \pi ^3}-\frac{13 {g}^2 {\langle\bar{q}q\rangle}^2   {\langle\bar{s}g\sigma Gs\rangle} {m_s}\delta (s)}{31104 \pi ^3}-\frac{17 {g}^2 \langle g^3G^{3}\rangle {\langle\bar{q}q\rangle}^2  \delta (s)}{11943936 \pi ^5}\\
    -&\frac{11 {g}^2 {\langle\bar{q}q\rangle}   {\langle\bar{s}s\rangle}^3\delta (s)}{2916 \pi }-\frac{11 {g}^2 {\langle\bar{q}q\rangle}^3   {\langle\bar{s}s\rangle}\delta (s)}{5832 \pi }-\frac{7 {g}^2 {\langle\bar{q}q\rangle}^2   {\langle\bar{s}s\rangle}^2\delta (s)}{11664 \pi }\\
    -&\frac{13 {\langle\bar{q}q\rangle} {\langle\bar{q}g\sigma Gq\rangle}   {\langle\bar{s}s\rangle} {m_s}\delta (s)}{96 \pi }-\frac{{\langle\bar{q}g\sigma Gq\rangle}   {\langle\bar{s}s\rangle}^2 {m_s}\delta (s)}{288 \pi }-\frac{19 {\langle\bar{q}q\rangle}^2   {\langle\bar{s}g\sigma Gs\rangle} {m_s}\delta (s)}{288 \pi }\\
    -&\frac{17 {\langle\bar{q}q\rangle}   {\langle\bar{s}s\rangle} {\langle\bar{s}g\sigma Gs\rangle} {m_s}\delta (s)}{864 \pi }+\frac{\langle g^3G^{3}\rangle {\langle\bar{q}q\rangle}   {\langle\bar{s}s\rangle}\delta (s)}{9216 \pi ^3}-\frac{\langle g^3G^{3}\rangle {\langle\bar{q}q\rangle}^2  \delta (s)}{27648 \pi ^3}\\
    +&\frac{\langle g^3G^{3}\rangle {\langle\bar{q}g\sigma Gq\rangle}   {m_s}\delta (s)}{49152 \pi ^5}+\frac{133 {\langle g^{2}G^{2}\rangle}^2   {\langle\bar{s}s\rangle} {m_s}\delta (s)}{5308416 \pi ^5}+\frac{73 {\langle g^{2}G^{2}\rangle}   {\langle\bar{s}s\rangle} {\langle\bar{s}g\sigma Gs\rangle}\delta (s)}{55296 \pi ^3}\\
    -&\frac{5 {g}^4   {\langle\bar{s}s\rangle}^4\delta (s)}{279936 \pi ^3}-\frac{13 {g}^2   {\langle\bar{s}s\rangle}^2 {\langle\bar{s}g\sigma Gs\rangle} {m_s}\delta (s)}{124416 \pi ^3}-\frac{31 {g}^2 \langle g^3G^{3}\rangle   {\langle\bar{s}s\rangle}^2\delta (s)}{11943936 \pi ^5}\\
    -&\frac{13 {g}^2   {\langle\bar{s}s\rangle}^4\delta (s)}{23328 \pi }+\frac{13   {\langle\bar{s}s\rangle}^2 {\langle\bar{s}g\sigma Gs\rangle} {m_s}\delta (s)}{864 \pi }-\frac{\langle g^3G^{3}\rangle   {\langle\bar{s}s\rangle}^2\delta (s)}{13824 \pi ^3}-\frac{17 \langle g^3G^{3}\rangle   {\langle\bar{s}g\sigma Gs\rangle} {m_s}\delta (s)}{1769472 \pi ^5}\, .\\
  \end{aligned}
  \label{rhop}
\end{equation}
\end{widetext}

\begin{widetext}
\begin{equation}
  \begin{aligned}
    \rho_{I}^{\text{OPE}}(s)=&\frac{3 m_s s^5}{5734400 \pi ^8}-\frac{\langle\bar{s}s\rangle s^4}{49152 \pi ^6}-\frac{\langle\bar{q}q\rangle s^4}{110592 \pi ^6}+\frac{7 \langle\bar{q}g\sigma Gq\rangle s^3}{92160 \pi ^6}+\frac{47 \langle\bar{s}g\sigma Gs\rangle s^3}{276480 \pi ^6}\\
    +&\frac{79 \langle g^{2}G^{2}\rangle m_s s^3}{8847360 \pi ^8}+\frac{23 \langle\bar{s}g\sigma Gs\rangle m_s^2 s^2}{36864 \pi ^6}+\frac{\langle\bar{q}q\rangle^2 m_s s^2}{768 \pi ^4}-\frac{\langle\bar{s}s\rangle^2 m_s s^2}{128 \pi ^4}\\
    +&\frac{13 \langle\bar{q}q\rangle \langle\bar{s}s\rangle m_s s^2}{768 \pi ^4}+\frac{13 g^{2} \langle\bar{q}q\rangle^2 m_s s^2}{165888 \pi ^6}-\frac{7  g^{2} \langle\bar{s}s\rangle^2 m_s s^2}{995328 \pi ^6}-\frac{799 \langle g^3G^{3}\rangle m_s s^2}{21233664 \pi ^8}\\
    -&\frac{25 \langle g^{2}G^{2}\rangle \langle\bar{q}q\rangle s^2}{442368 \pi ^6}-\frac{71 \langle g^{2}G^{2}\rangle \langle\bar{s}s\rangle s^2}{442368 \pi ^6}+\frac{\langle\bar{s}s\rangle^3 s}{216 \pi ^2}-\frac{17 \langle\bar{q}q\rangle \langle\bar{s}s\rangle^2 s}{216 \pi ^2}\\
    -&\frac{7 \langle\bar{q}q\rangle^2 \langle\bar{s}s\rangle s}{864 \pi ^2}+\frac{ g^{2} \langle\bar{s}s\rangle^3 s}{11664 \pi ^4}-\frac{g^{2} \langle\bar{q}q\rangle^2 \langle\bar{s}s\rangle s}{2592 \pi ^4}-\frac{5 g^{2}  \langle\bar{q}q\rangle^3 s}{3888 \pi ^4}+\frac{7 g^{2} \langle\bar{q}q\rangle \langle\bar{s}s\rangle^2 s}{3888 \pi ^4}\\
    +&\frac{\langle g^3G^{3}\rangle \langle\bar{q}q\rangle s}{6144 \pi ^6}+\frac{67 \langle g^3G^{3}\rangle \langle\bar{s}s\rangle s}{248832 \pi ^6}+\frac{341 \langle g^{2}G^{2}\rangle \langle\bar{q}g\sigma Gq\rangle s}{5308416 \pi ^6}\\
    +&\frac{1067 \langle g^{2}G^{2}\rangle \langle\bar{s}g\sigma Gs\rangle s}{5308416 \pi ^6}+\frac{29 \langle\bar{s}s\rangle \langle\bar{s}g\sigma Gs\rangle m_s s}{1728 \pi ^4}-\frac{85 \langle\bar{q}q\rangle \langle\bar{s}g\sigma Gs\rangle m_s s}{2304 \pi ^4}\\
    -&\frac{41 \langle\bar{q}q\rangle \langle\bar{q}g\sigma Gq\rangle m_s s}{4608 \pi ^4}-\frac{149 \langle\bar{s}s\rangle \langle\bar{q}g\sigma Gq\rangle m_s s}{6912 \pi ^4}\\
    +&\frac{5 {\langle g^{2}G^{2}\rangle} {g}^2 {\langle\bar{q}q\rangle}^2 {m_s}}{746496 \pi ^6}+\frac{149 {\langle g^{2}G^{2}\rangle} {\langle\bar{q}q\rangle} {\langle\bar{s}s\rangle} {m_s}}{27648 \pi ^4}-\frac{7 {\langle g^{2}G^{2}\rangle} {\langle\bar{q}q\rangle}^2 {m_s}}{6912 \pi ^4}\\
    +&\frac{{g}^2 {\langle\bar{q}g\sigma Gq\rangle} {\langle\bar{s}s\rangle}^2}{5184 \pi ^4}+\frac{{g}^2 {\langle\bar{q}q\rangle}^2 {\langle\bar{s}g\sigma Gs\rangle}}{3456 \pi ^4}+\frac{5 {\langle\bar{q}g\sigma Gq\rangle} {\langle\bar{s}s\rangle}^2}{144 \pi ^2}\\
    +&\frac{{\langle\bar{q}q\rangle} {\langle\bar{q}g\sigma Gq\rangle} {\langle\bar{s}s\rangle}}{288 \pi ^2}+\frac{{\langle\bar{q}q\rangle}^2 {\langle\bar{s}g\sigma Gs\rangle}}{576 \pi ^2}+\frac{5 {\langle\bar{q}q\rangle} {\langle\bar{s}s\rangle} {\langle\bar{s}g\sigma Gs\rangle}}{72 \pi ^2}\\
    +&\frac{{\langle\bar{q}g\sigma Gq\rangle}^2 {m_s}}{3072 \pi ^4}+\frac{3 {\langle\bar{q}g\sigma Gq\rangle} {\langle\bar{s}g\sigma Gs\rangle} {m_s}}{256 \pi ^4}-\frac{5 {\langle g^{2}G^{2}\rangle}^2 {\langle\bar{q}q\rangle}}{2654208 \pi ^6}+\frac{{g}^2 {\langle\bar{q}q\rangle}^2 {\langle\bar{q}g\sigma Gq\rangle}}{15552 \pi ^4}\\
    +&\frac{5 {\langle g^3G^{3}\rangle} {\langle\bar{q}g\sigma Gq\rangle}}{294912 \pi ^6}-\frac{29 {\langle g^{2}G^{2}\rangle}^2 {\langle\bar{s}s\rangle}}{5308416 \pi ^6}+\frac{47 {\langle g^{2}G^{2}\rangle} {g}^2 {\langle\bar{s}s\rangle}^2 {m_s}}{5971968 \pi ^6}\\
    -&\frac{25 {\langle g^{2}G^{2}\rangle} {\langle\bar{s}s\rangle}^2 {m_s}}{27648 \pi ^4}+\frac{{g}^2 {\langle\bar{s}s\rangle}^2 {\langle\bar{s}g\sigma Gs\rangle}}{3456 \pi ^4}-\frac{{\langle\bar{s}s\rangle}^2 {\langle\bar{s}g\sigma Gs\rangle}}{192 \pi ^2}\\
    -&\frac{3 {\langle\bar{s}g\sigma Gs\rangle}^2 {m_s}}{1024 \pi ^4}+\frac{7 {\langle g^3G^{3}\rangle} {\langle\bar{s}g\sigma Gs\rangle}}{294912 \pi ^6}-\frac{{\langle g^{2}G^{2}\rangle} {g}^2 {\langle\bar{q}q\rangle}^3 \delta (s)}{139968 \pi ^3}\\
    -&\frac{{\langle g^{2}G^{2}\rangle} {g}^2 {\langle\bar{q}q\rangle}^2  {\langle\bar{s}s\rangle}\delta (s)}{69984 \pi ^3}+\frac{5 {\langle g^{2}G^{2}\rangle} {\langle\bar{q}q\rangle}  {\langle\bar{s}g\sigma Gs\rangle} {m_s}\delta (s)}{6912 \pi ^3}-\frac{17 {\langle g^{2}G^{2}\rangle} {\langle\bar{q}q\rangle} {\langle\bar{q}g\sigma Gq\rangle} {m_s}\delta (s)}{27648 \pi ^3}\\
    +&\frac{5 {\langle g^{2}G^{2}\rangle} {\langle\bar{q}g\sigma Gq\rangle}  {\langle\bar{s}s\rangle} {m_s}\delta (s)}{6912 \pi ^3}-\frac{17 {\langle g^{2}G^{2}\rangle} {\langle\bar{q}q\rangle}^2   {\langle\bar{s}s\rangle}}{10368 \pi }+\frac{5 {\langle g^{2}G^{2}\rangle} {\langle\bar{q}q\rangle}   {\langle\bar{s}s\rangle}^2\delta (s)}{2592 \pi }-\frac{{g}^2 \langle g^3G^{3}\rangle {\langle\bar{q}q\rangle}^2   {m_s}\delta (s)}{1492992 \pi ^5}\\
    -&\frac{{g}^2 {\langle\bar{q}q\rangle}^3   {\langle\bar{s}s\rangle} {m_s}\delta (s)}{486 \pi }+\frac{{g}^2 {\langle\bar{q}q\rangle}^2   {\langle\bar{s}s\rangle}^2 {m_s}\delta (s)}{2592 \pi }-\frac{{g}^2 {\langle\bar{q}q\rangle}   {\langle\bar{s}s\rangle}^3 {m_s}\delta (s)}{972 \pi }+\frac{{\langle\bar{q}q\rangle}   {\langle\bar{s}g\sigma Gs\rangle}^2\delta (s)}{192 \pi }\\
    -&\frac{{\langle\bar{q}q\rangle} {\langle\bar{q}g\sigma Gq\rangle}   {\langle\bar{s}g\sigma Gs\rangle}\delta (s)}{384 \pi }+\frac{{\langle\bar{q}g\sigma Gq\rangle}   {\langle\bar{s}s\rangle} {\langle\bar{s}g\sigma Gs\rangle}\delta (s)}{96 \pi }-\frac{{\langle\bar{q}g\sigma Gq\rangle}^2   {\langle\bar{s}s\rangle}\delta (s)}{768 \pi }\\
    +&\frac{\langle g^3G^{3}\rangle {\langle\bar{q}q\rangle}   {\langle\bar{s}s\rangle} {m_s}\delta (s)}{13824 \pi ^3}-\frac{10\pi  {\langle\bar{q}q\rangle}^2   {\langle\bar{s}s\rangle}^2 {m_s}\delta (s)}{27} +\frac{7\pi  {\langle\bar{q}q\rangle}   {\langle\bar{s}s\rangle}^3 {m_s}\delta (s)}{54} -\frac{{\langle g^{2}G^{2}\rangle} {g}^2   {\langle\bar{s}s\rangle}^3\delta (s)}{93312 \pi ^3}\\
    +&\frac{11 {\langle g^{2}G^{2}\rangle}   {\langle\bar{s}s\rangle} {\langle\bar{s}g\sigma Gs\rangle} {m_s}\delta (s)}{27648 \pi ^3}+\frac{{\langle g^{2}G^{2}\rangle}   {\langle\bar{s}s\rangle}^3\delta (s)}{5184 \pi }-\frac{{g}^2 \langle g^3G^{3}\rangle   {\langle\bar{s}s\rangle}^2 {m_s}\delta (s)}{2985984 \pi ^5}\\
    +&\frac{\langle g^3G^{3}\rangle   {\langle\bar{s}s\rangle}^2 {m_s}\delta (s)}{13824 \pi ^3}\, .
  \end{aligned}
  \label{rhoi}
\end{equation}
\end{widetext}

\textcolor{black}{}

%\bibliographystyle{paper}
%\bibliography{ref}

\end{document}